# k-dependent proximity-induced modulation of spin-orbit interaction in MoSe$_2$ interfaced with amorphous Pb

*Fatima Alarab* [1], *Ján Minár* [2], *Procopios Constantinou* [1], *Dhani Nafday*[3], *Thorsten Schmitt* [1], *Xiaoqiang Wang* [1] *and Vladimir N. Strocov* [1]

[1]Swiss Light Source, Paul Scherrer Institute, 5232 Villigen-PSI, Switzerland
[2]New Technologies Research Centre, University of West Bohemia, 301 00 Plzeň, Czech Republic
[3]Asia Pacific Center for Theoretical Physics, Pohang, Gyeongbuk, 37673, South Korea

## Abstract

The ability to modulate the spin-orbit (SO) interaction is crucial for engineering a wide range of spintronics-based quantum devices, extending from state-of-the-art data storage to materials for quantum computing. The use of proximity-induced effects for this purpose may become the mainstream approach, whereas their experimental verification using angle-resolved photoelectron spectroscopy (ARPES) has so far been elusive. Here, using the advantages of soft-X-ray ARPES on its probing depth and intrinsic resolution in three-dimensional momentum **k**, we identify a distinct modulation of the SO interaction in a van der Waals semiconductor (MoSe$_2$) proximitized to a high-Z metal (Pb), and measure its variation through the **k**-space. The strong SO field from Pb boosts the SO splitting by up to 30% at the H-point of the bulk Brillouin zone, the spin-orbit hotspot of MoSe$_2$. Tunability of the splitting via the Pb thickness allows its tailoring to particular applications in emerging quantum devices.

# Introduction

Spintronics is based on novel operational principles that utilise the electron spin degree of freedom to process information, which offers a greater diversity of functionality with respect to conventional electronics [1-3]. Spintronic devices are also expected to be the leading contenders of next generation nanoelectronics devices, promising reduced power consumption, increased memory density and processing capabilities. Typically, in such spintronic devices, the spin polarisation is controlled either by magnetic layers, or via spin-orbit (SO) coupling [4].

SO coupling is known to play a key role in a variety of the novel functionalities that underpin the materials used for spintronics. Such functionalities, strongly depending on the SO coupling strength, are behind the anomalous Hall effect [5], SO torques [3], anisotropic magnetoresistance [6] and spin relaxation [7] in magnetic materials, as well as in the interconnection between nonequilibrium charge and spin currents in non-magnetic semiconductors [8,9], and the Rashba-Edelstein effect (inverse Rashba-Edelstein effect) in topological insulator materials [10-14]. Furthermore, strong SO coupling is the key parameter to drive, for example, high-dimensional topological features such as robust surface states in topological insulators [15,16] as well as three-dimensional nodal lines and chains in Dirac and Weyl semimetals [17-22]. In addition to the spintronic technologies, tunable SO coupling is vital in engineering novel exotic states of matter such as Majorana fermions, envisaged for qubits in quantum computers, which are based on interfacing superconductors with strong SO coupling semiconductors [23-28]. The SO coupling strength is critical here for the topological protection of Majorana modes against the loss of coherence due to other low-energy excitations [24,29,30]. Thus, finding practical and controllable methods to tune the SO strength is critical for the development of new functional materials that can be used in the next-generation of quantum devices.

In principle, it is possible to increase the SO strength of a material by either doping or interfacing it with high-$Z$ elements. The latter is based on the hybridization of the host wavefunctions with those of the heavy atoms delivering a large SO field, which results in a proximity-induced enhancement of the SO splitting ($\Delta_{SO}$). In this way, $\Delta_{SO}$ strongly depends on the electronic charge distribution and on hybridization effects in the interface region. So far, most attempts to study the proximity effects with ARPES have focused on graphene grown on diverse substrates of 5$d$ metals such as Au, Ir and Pt [31-37]. In these cases, only an increase of the Rashba spin splitting due to changes in the charge distribution at the interface and a band gap formation due to electronic hybridization with the substrate's electron states were detected, without a notable proximity-induced intrinsic SO splitting. Placing graphene onto transition-metal dichalcogenides (TMDCs) would be another strategy to control the SO coupling in graphene. Using magnetotransport measurements [38] and scanning tunneling spectroscopy imaging [39], both Rashba and valley Zeeman terms in the proximity-induced SO coupling were identified, with a magnitude of 7-16 meV and 1-3 meV, respectively, and the origin of this phenomenon was traced back to intrinsic defects in the substrates. Yet, the associated band splitting has so far escaped a direct **k**-resolved investigation with ARPES, mainly because the small splitting magnitude is comparable with the experimental resolution limit. In any case, the gapless nature of graphene makes it hardly usable for transistor applications.

In this regard, the diverse family of TMDCs, composed of weakly van der Waals (vdW) bonded atomic layers, can be more attractive [40-43]. Having a finite semiconducting band gap and diverse band structure patterns, these materials are promising candidates for applications in microelectronic and spintronic devices. One such TMDC is MoSe$_2$, an indirect band gap semiconductor crystallising in its thermodynamically stable 2H-phase shown in **Fig. 1(a,b)**. Its trigonal single-layer structure forbids inversion symmetry, thereby allowing intrinsic SO coupling with an energy splitting of up to ~200 meV in its valence bands at the H-point (see **Fig. 1c**). Since this splitting is relatively large, MoSe$_2$ is an ideal host to explore the proximity-induced enhancement of the SO splitting by interfacing the surface with a high-Z element.

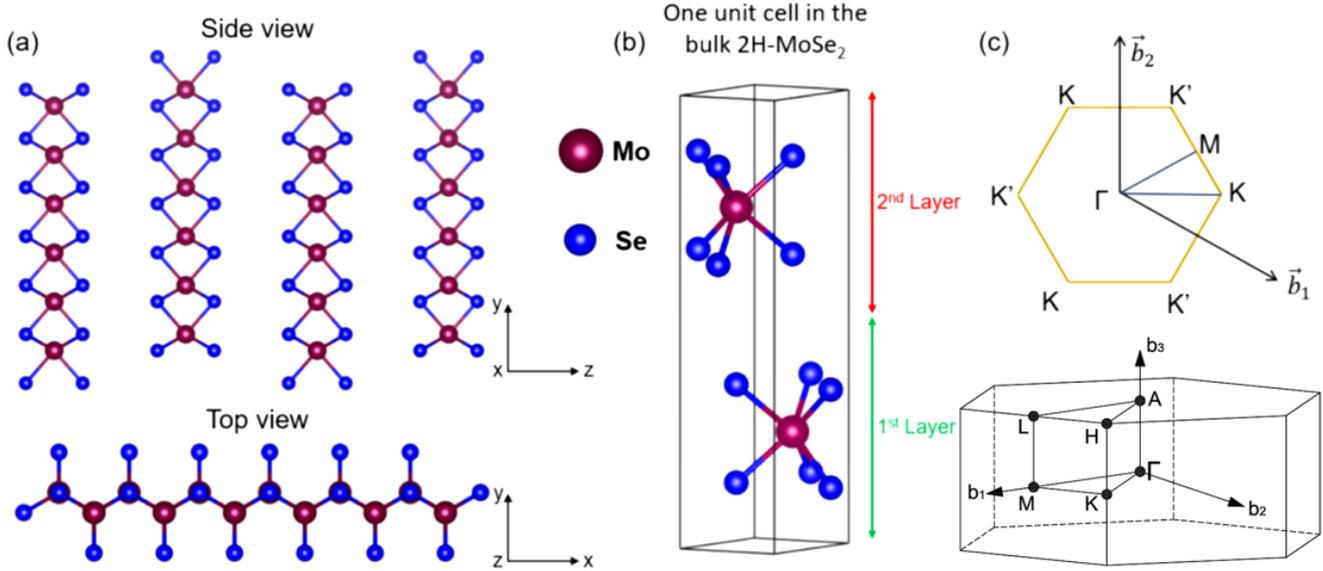

Fig. 1: **Crystallography of 2H-MoSe$_2$.** (a) Side- and top-view of MoSe$_2$ crystal structure in its prismatic 2H-phase. (b) One unit cell in the bulk 2H-MoSe$_2$. (c) Corresponding 2D (top) and 3D (bottom) Brillouin zones of MoSe$_2$ showing the high-symmetry points.

Here, we observe and quantify the anticipated proximity effect induced by the strong SO field originating from a Pb overlayer interfaced with MoSe$_2$. An enhancement in the SO splitting, $\Delta_{SO}$, in certain **k**-space regions was observed, which can be tuned by controlling the surface coverage of Pb. We explored the texture of $\Delta_{SO}$ through the three-dimensional (3D) band structure of MoSe$_2$ using soft-X-ray ARPES with photon energies (*hν*) towards 1 keV, where the increase of the photoelectron escape depth compared to the conventional VUV energy range increases the intrinsic definition ($\Delta k_z$) of the out-of-plane electron momentum $k_z$ [44-46]. A $\Delta_{SO}$ magnitude of 190 meV found in the H-point of the 3D Brillouin zone (3D-BZ) of MoSe$_2$ is sufficiently small to be sensitive to the external SO field and at the same time large enough compared to the experimental energy resolution. To identify the proximity effect, we monitored the evolution of $\Delta_{SO}$ in this hotspot upon low-temperature deposition of Pb. The amorphous character of this film eliminated potential effects of **k**-space mismatch of the wavefunctions across the interface, and the large probing depth of soft-X-ray ARPES was essential to penetrate through the Pb film to MoSe$_2$. This finding paves the way towards fabrication of TMDC-based spintronic devices where the SO coupling strength is tuned via proximity effect with high-Z atoms. The deposition of their amorphous films will significantly simplify the fabrication of such devices compared to the advanced epitaxial deposition techniques. Our findings can naturally extend to the TMDC-based materials where the strong SO field is provided by intercalation of high-Z atoms.

## Results and discussion

### SO splitting hotspots in the valence band of MoSe$_2$

The electronic structure of the TMDC systems critically depends on the number of layers, going from the bulk to the monolayer (ML) limit [47]. The corresponding evolution of the electronic structure is driven by a modification in the hybridization between the $p_z$ and $d$ orbitals originating from the Se and Mo atoms respectively. This is responsible for the interlayer interaction, quantum confinement, and, for systems with odd numbers of MLs, denial of the inversion symmetry resulting in Rashba-type SO splitting at the Γ-point [20,47-49].

Most of the recent experimental and theoretical work on MoSe$_2$ [20,40,50-59] focused on the valence band (VB) dispersions at the K-point which is identified in the 3D-BZ of **Fig. 1 (c)**. They show that the band splitting ($\Delta_{bs}$) at this point arises, in the bulk limit, from the interlayer interaction, with a smaller contribution of the intrinsic SO coupling and, in the ML limit, from the strong SO interaction caused by the absence of the inversion centre.

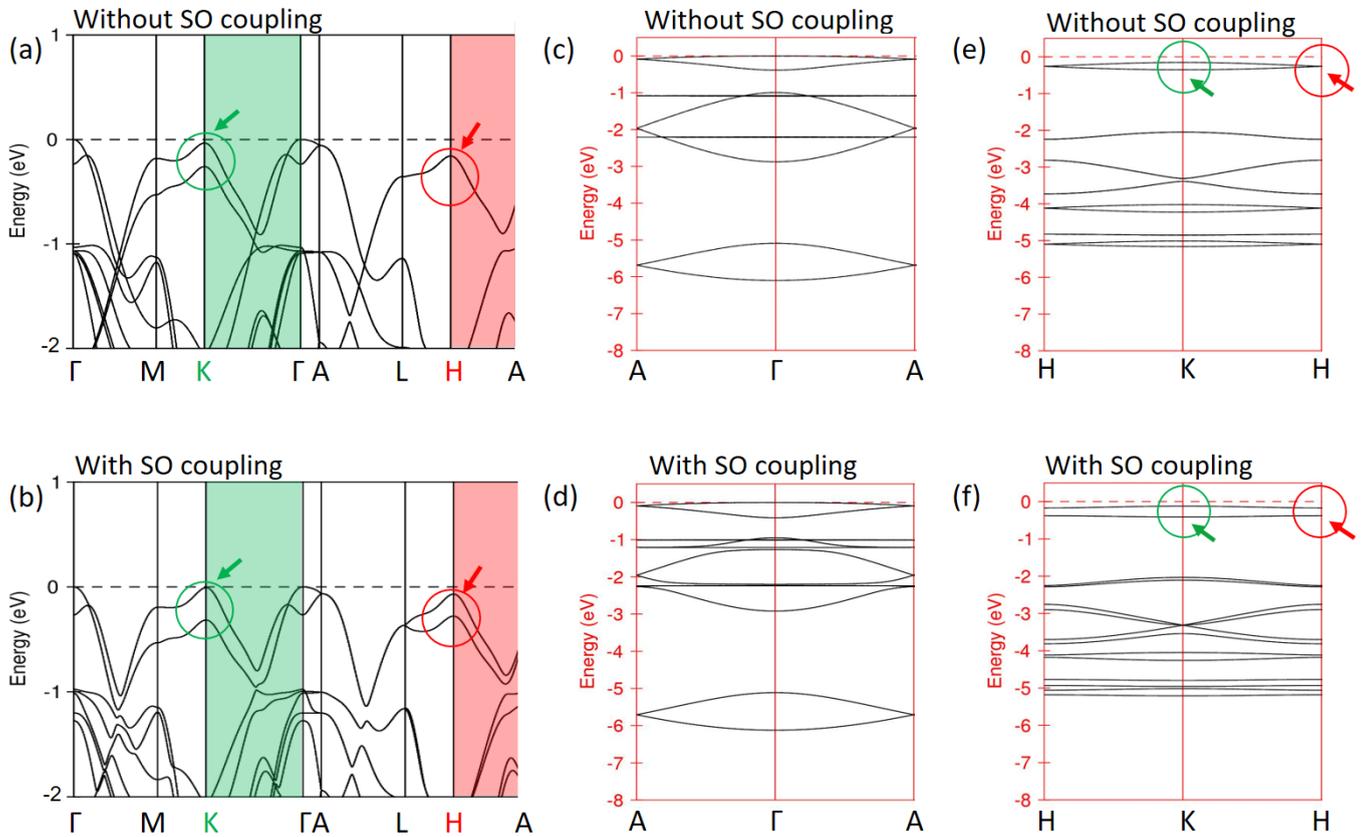

Fig. 2: **Theoretical band structure of MoSe$_2$.** (a,b) DFT calculations for the electronic band structures of bulk MoSe$_2$ without and with SO coupling, respectively. The SO-coupling hotspot is located in the H-point. DFT calculations showing the dispersion of the bands along the out of plane high-symmetry directions (c,d) AΓA and (e,f) HKH of bulk MoSe$_2$ without and with SO coupling. The green/red circles and arrows highlight the valence band splitting at the K/H points. The dashed lines correspond to the Fermi level ($E_F$=0 eV).

This picture is confirmed by our DFT calculations for bulk MoSe$_2$, along the ΓK- and AH-directions, without and with the SO coupling shown in **Fig. 2 (a,b)**, respectively. At the K-point, $\Delta_{bs}$ near the Fermi level $E_F$ increases only slightly, when the SO coupling is added. This splitting is therefore controlled mostly by the interlayer interaction with only a weak contribution of SO coupling. At the H-point, which differs from K only by its $k_z$, the bands near $E_F$ split only when the SO coupling is activated. Therefore, in contrast to the sister K-point, $\Delta_{bs}$ at this point results solely from the strong SO interaction. This makes the H-point the ideal hotspot for analysis of the SO interaction and its modulation driven by the proximity effect. A detailed description of the electronic band structure calculations in the out of plane directions, without and with SO coupling is shown in **Fig. 2 (c-e)** along AΓA and HKH. Our results show a number of nearly perfect dispersionless bands located at about -1.1 and -2.3 eV along AΓA and at around -5 eV along HKH, and are splitted due to the SO interaction. These bands can be attributed to Mo 4*d* states which therefore explains their two-dimensional character. The top of the valence band, as well as the band with largest binding energies reveal a significant out of plane momentum dependence. One can assume that these bands are predominantly of $p_z$ orbital character, which is strongly influenced by the sandwich layered structure in the crystal (see **Fig. 1 (b)**). Most importantly are the features observed along the HKH direction. Here, $MoSe_2$ shows a distinct SO splitting at the H point in addition to a clear three-dimensional band character. To experimentally define the different hot spots where the band splitting is observed, these data will be cross compared with our photoemission measurements.

## 3D electronic band dispersion of bulk $MoSe_2$

To identify the SO-coupling hotspots in the ARPES data, we will first establish the experimental 3D band structure of bare $MoSe_2$. **Fig. 3(a)** shows the out-of-plane $k_{xz}$ slices of ARPES spectral intensity taken at two constant binding energies, $E_b$; one at 0 eV (which corresponds to the valence band maximum (VBM) at the Γ-point) and the other at -0.65 eV. A dispersive periodic pattern is observed along the out-of-plane $k_z$ axis which originates from the 3D electronic structure and allows us to determine the position of the Γ points. This is important, as a 3D electronic structure is required to disentangle the bands at the K and H-points since they only differ slightly in $k_z$. In **Fig. 3 (b)**, an additional sequence of four in-plane $k_{xy}$ slices is shown at different $E_b$; $E_b$=0 eV, -0.3 eV, -0.65 eV, and -1.85 eV. At $E_b$=-0.65 eV, the iso-$E_b$ map, aligned along the ΓK-direction, shows a hexagonal pattern of closed contours around the Γ and K-points; the aforementioned interlayer splitting manifests itself as splitting of the contours around the K-point. This effect is in agreement with our calculations in **Fig. 2 (b)** as well as with the previous theoretical and experimental data [53,59].

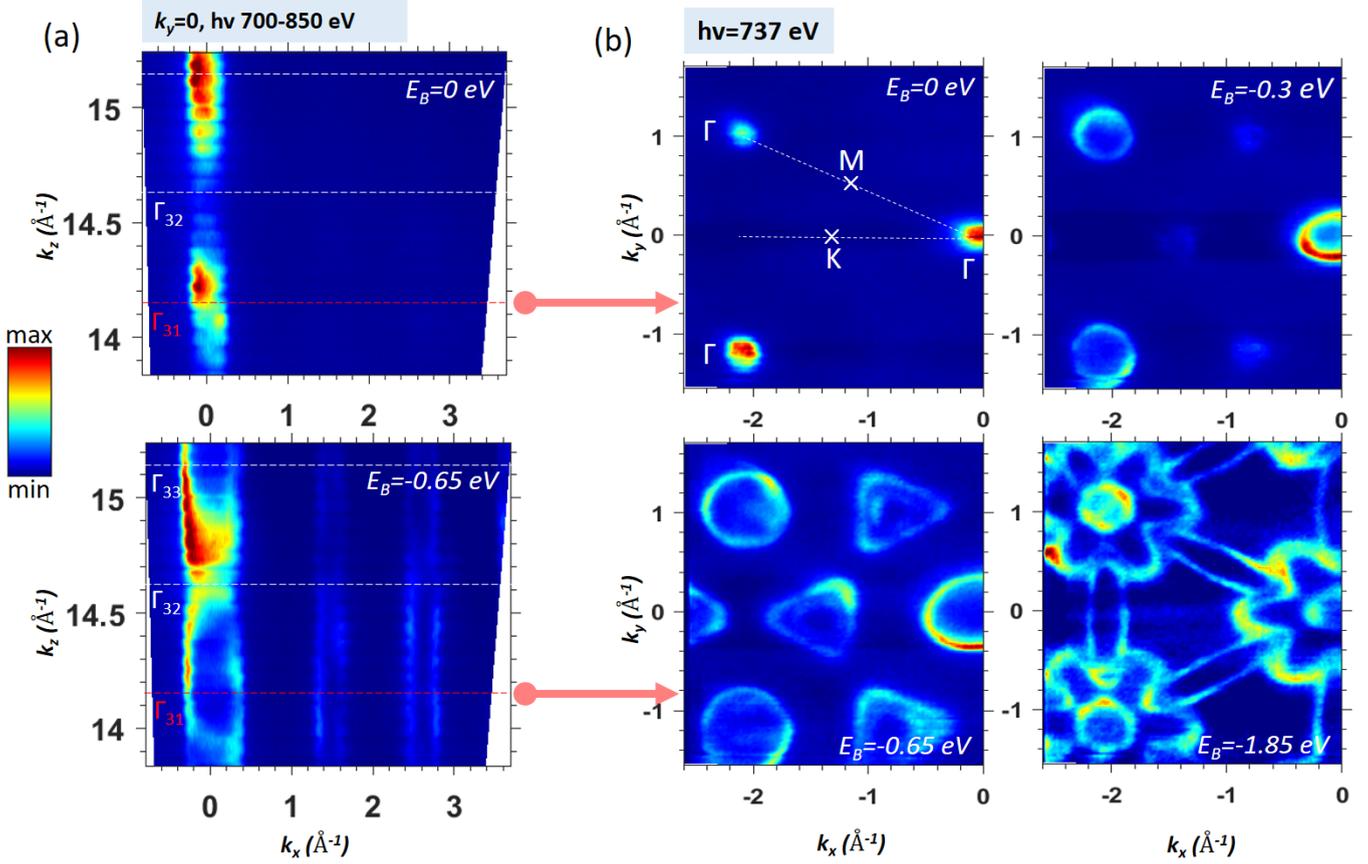

Fig. 3: **Experimental band structure of bare MoSe$_2$.** (a) Out-of-plane momentum $k_{xz}$ cuts of ARPES intensity in the $h\nu$ range 700-850 eV, taken at $k_y$=0 and for E$_b$=0 eV, and -0.65 eV, showing the periodicity along $k_z$. The conversion from $h\nu$ to $k_z$ was carried out within the free-electron approximation for the final state with an inner potential of 14 eV. (b) In-plane momentum $k_{xy}$ cuts of ARPES intensity taken at $E_b$=0 eV, -0.3 eV, -0.65 eV and -1.85 eV. These measurements were performed at $h\nu$=737 eV, which corresponds to the central ($k_z$=0) plane of the bulk 3D-BZ (see **Fig. 1 (c)**). The high-symmetry points are indicated showing the hexagonal structure of the lattice system as well as the trigonal warping effect near K. $E_b$=0 eV corresponds to the valence band maxima at the Γ-point.

Directly reflecting the out-of-plane dispersion of the valence states, **Fig. 4 (a)** shows the ARPES intensity maps as a function of $E_b$ and $k_z$ measured at $k_x$=0 Å$^{-1}$ (along ΓA) and 1.6 Å$^{-1}$ (along KH) in the $h\nu$ range from 700 eV to 850 eV. We note that the VB of MoSe$_2$ is formed by the Mo 4$d$ and Se 4$p$ states being strongly hybridised through the 3D-BZ [60-62]. The experimental maps show dispersive bands where the two low-$E_b$ bands are formed by the bonding Se 4$p_z$ and antibonding Se 4$p_z$* out-of-plane orbitals overlapping through the vdW gap, and the one at the VB maximum by the Mo 4$d_{xy}$ orbitals whose interlayer cross-talk is mediated by their hybridization with the Se 4$p_z$* orbitals [60]. The dispersion of these bands shows a good agreement with our previously discussed DFT calculations, and is going through the indicated sequence of the high-symmetry points Γ to A and K to H highlighted in the two panels. The periodicity of these dispersions is twice the out-of-plane size of the 3D-BZ, which is characteristic of the materials with a non-symmorphic space group [44,63,64]. The $h\nu$ value 737 eV, where the in-plane iso-$E_b$ maps in **Fig. 3 (b)** were recorded, corresponds to the Γ-point along the ΓA-direction and, due to small variations of $k_z$ with $k_{xy}$ in the soft-X-ray range, almost to the K-point along KH.

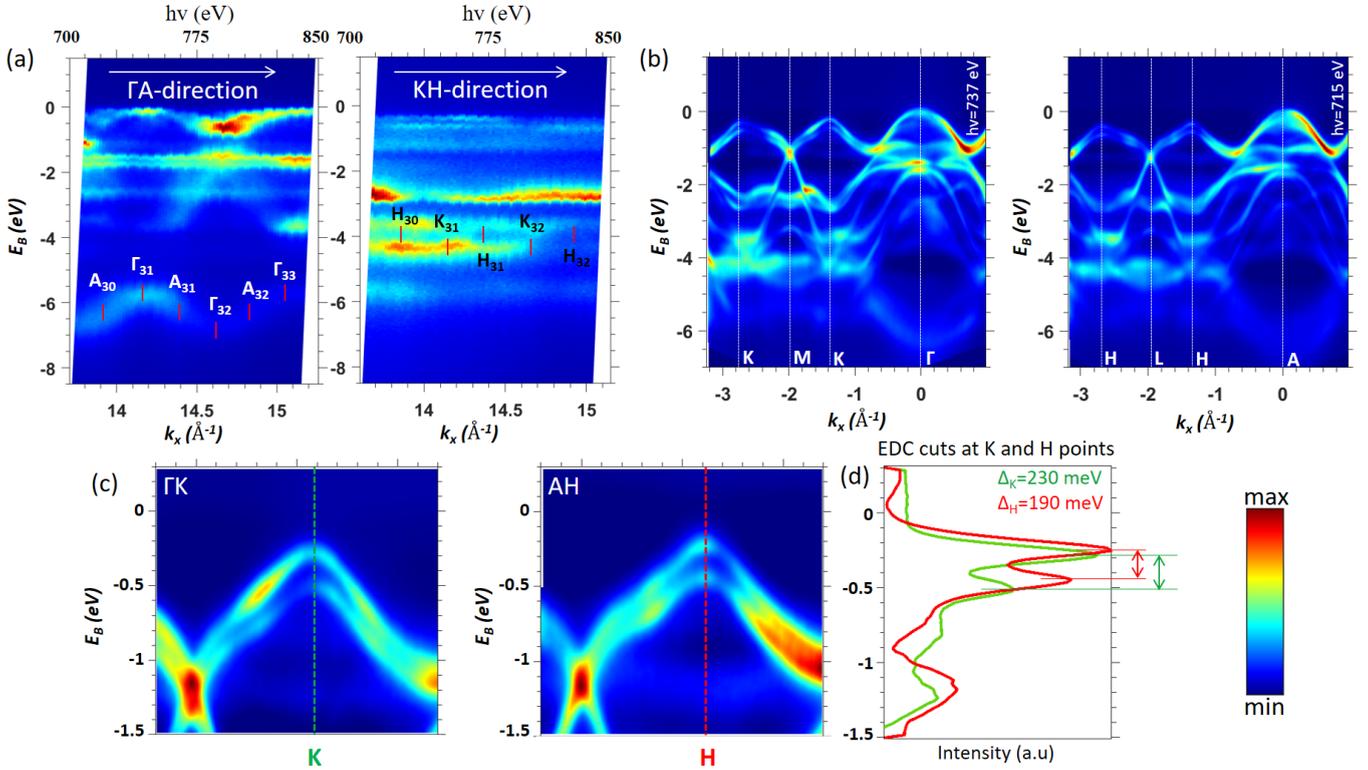

Fig. 4: **Intrinsic SO splitting in bare MoSe$_2$.** (a) Out-of-plane ARPES intensity map as a function $k_z$ in the $hv$ range 700-850 eV in the ΓA (left panel) and KH (right panel) directions. The green arrows indicate the out-of-plane dispersive bands. (b) ARPES intensity images measured at $hv$=737 eV and 715 eV, respectively, where **k** runs through the ΓKMK (left panel) and AHLH (right panel) directions of the 3D-BZ, showing $\Delta_{bs}$ at the K and H-points. (c) Zoom-in of (b) near the K and H-points. (d) EDCs in the K (green line) and H (red) points, showing the corresponding $\Delta_{bs}$ values. The splitting at the K-point represents mostly the interlayer interaction, and at the H-point hotspot mostly $\Delta_{SO}$.

We will now focus on $\Delta_{bs}$ at the K and H-points, which is central for our analysis of the SO interaction. **Fig. 4 (b)** shows the ARPES intensity images measured at $hv$=790 eV, where **k** runs through the ΓKMK and AHLH-directions of the 3D-BZ (see **Fig. 1 (c)**). The experimental bands near the K and H-points clearly split near the VBM, which is most evident in the zoom-ins of **Fig. 4 (b)**. The Energy-Distribution Curves (EDCs), plotted in **Fig. 4 (c)** as the green and red lines for K and H respectively, quantify the corresponding splitting magnitudes as 230 meV and 190 meV. These figures are consistent with our DFT calculations including the SO coupling in **Fig. 2 (b)**; the splitting at K originates mainly from the interlayer interaction, whereas the splitting at H originates only from the SO interaction and is therefore the hotspot for monitoring any changes in the SO coupling. We note that the K and H-points differ only by their $k_z$, and the small intrinsic $\Delta k_z$ in the soft-X-ray energy range is quintessential to distinguish their fingerprints in the experimental spectra. Indeed, according to the TPP-2M formula [65,66], the photoelectron inelastic mean free path increases in our kinetic-energy range to ~15.7 Å. This gives $\Delta k_z$~0.06 Å$^{-1}$, which is sufficiently sharp relative to the $k_z$ separation of the K and H-points (0.464 Å$^{-1}$).

## Band structure evolution with Pb thickness

To study the proximity-induced SO interaction in MoS$_2$ we deposited Pb *in-situ* on its cleaved crystalline surface, and tracked the evolution of $\Delta_{bs}$ at the K and the H-points versus Pb layer thickness. Pb was deposited on the MoSe$_2$ surface in successive steps (see Methods) followed by X-ray photoemission spectroscopy (XPS) and ARPES measurements for each deposition. The XPS spectra of the Mo 3$d$ and Pb 4$f$ core-level peaks, **Fig. 5 (a)**, show that they stayed at the same $E_b$ without any noticeable lineshape changes, confirming the absence of chemical reactions between the Pb and MoSe$_2$ or formation of new chemical environments of the interfacial atoms. Furthermore, the ARPES data did not reveal any states additional to bare MoSe$_2$, **Fig. 6 (a,b)**, apart from the incoherent background coming from the amorphous Pb layer that piled up with the increase of its thickness, **Fig. 5 (c)**. Formation of homogeneous amorphous overlayers is essential for our study. As Pb does not wet the MoSe$_2$ surface, the depositions were carried out at 12 K. Furthermore, the absence of islands in the overlayer has been verified by a linear dependence of the Mo 3$d$ and Pb 4$f$ intensities as a function of Pb deposition dose, **Fig. 5 (a)**, and the correspondence between deposition time and overlayer thickness is presented in **Fig. 5 (b)** (for detailed analysis see Supplementary).

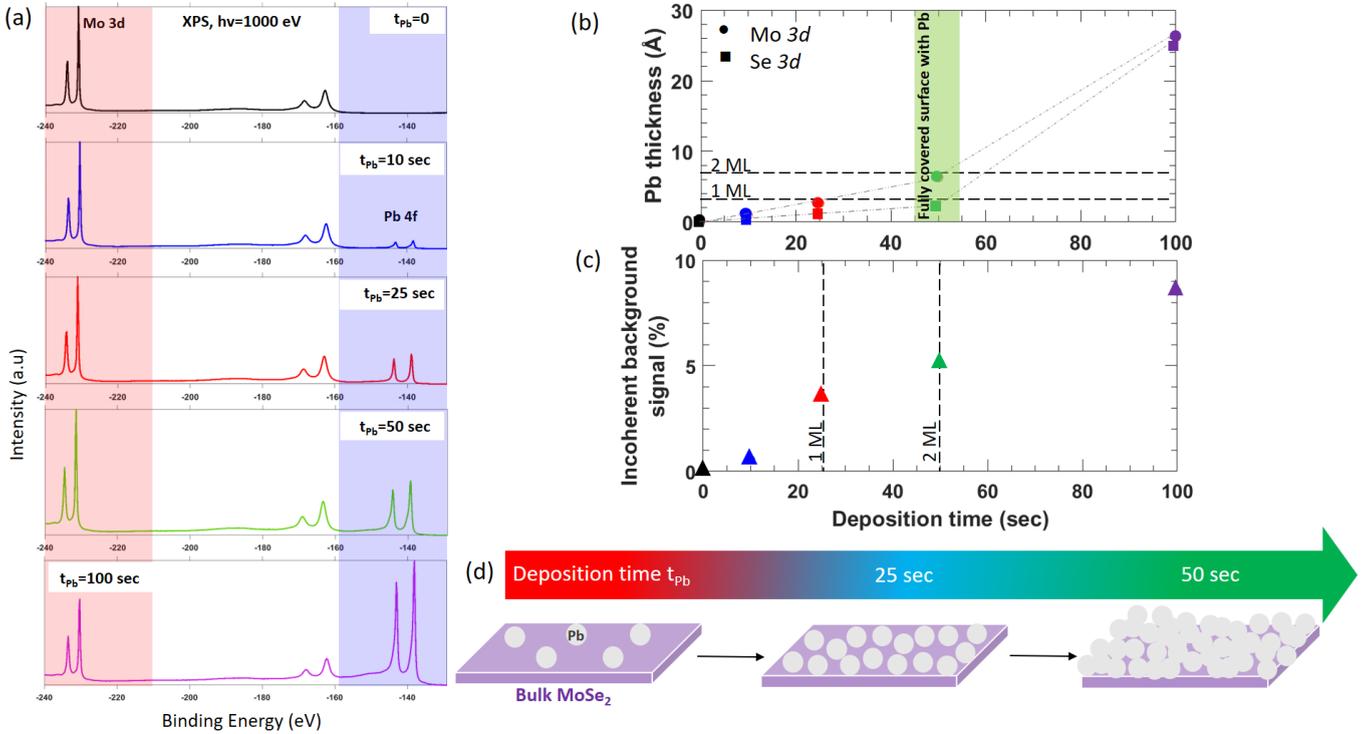

Fig. 5: **Growth of amorphous Pb film at the surface of MoSe$_2$ crystal.** (a) XPS overview recorded at $h\nu$=1000 eV showing the evolution of the Mo 3$d$ and Pb 4$f$ core levels through the series of Pb overlayer thicknesses. (b) Correlation between the deposition time and Pb thickness. The latter is determined by the intensity attenuation of the Mo 3$d$ and Se 3$d$ core levels upon Pb deposition. (c) Correlation between the deposition time and the incoherent background of our ARPES data. (d) Sketch showing the different distribution of Pb atoms on the MoSe$_2$ surface with deposition time.

**Fig. 6 (a,b)** shows our ARPES data around the K and H-points for MoSe$_2$ covered with 0.35 and 1 ML of Pb, respectively, corresponding to the zoom-ins for bare MoSe$_2$ in **Fig. 4 (c)**. The EDCs for the K and H-points, shown on the right in **Fig. 6 (c,d)**, represent $\Delta_{bs}$ at each point. Its evolution with the Pb layer's thickness from 0 to 9 MLs at the H$_{30}$, H$_{31}$, K$_{31}$, and K$_{32}$ points of the experimental $k_z$ dependence (**Fig. 4 (a)**) is shown in **Fig. 6 (e)**. The error bars of $\Delta_{bs}$ at each point was calculated from measurements over a series of six crystals at the same deposition stages using Student's $t$-distribution function [67]. These results show that at the K-points $\Delta_{bs}$ stays constant at ~230 meV independently of the Pb thickness. In contrast, at the H-points $\Delta_{bs}$ increases from 190 meV (for bare MoSe$_2$) to 230 and 250 meV upon depositing 0.35 and 1 ML of Pb respectively. It saturates at 270 meV once Pb forms two full MLs on the MoSe$_2$ surface, and does not increase upon further deposition of Pb.

The thickness of Pb overlayers is evidently a key parameter to control the $\Delta_{bs}$ magnitude. Upon the evaporation process, Pb atoms are randomly deposited at the surface of the substrate (**Fig. 5 (d)** left panel). The electronic orbitals of the nearest neighbouring Pb atoms communicate to form a local potential and generate an SO field, which is then transmitted to the substrate. By increasing progressively the thickness of the evaporated Pb, more atoms interact and induce a stronger SO field in the Pb layer (**Fig. 5 (d)** middle panel) until it reaches its maximum value when two full Pb MLs are formed (**Fig. 5 (d)** right panel). At this level, the modulated SO coupling induced by proximity at the interface remains at its peak, saturated value.

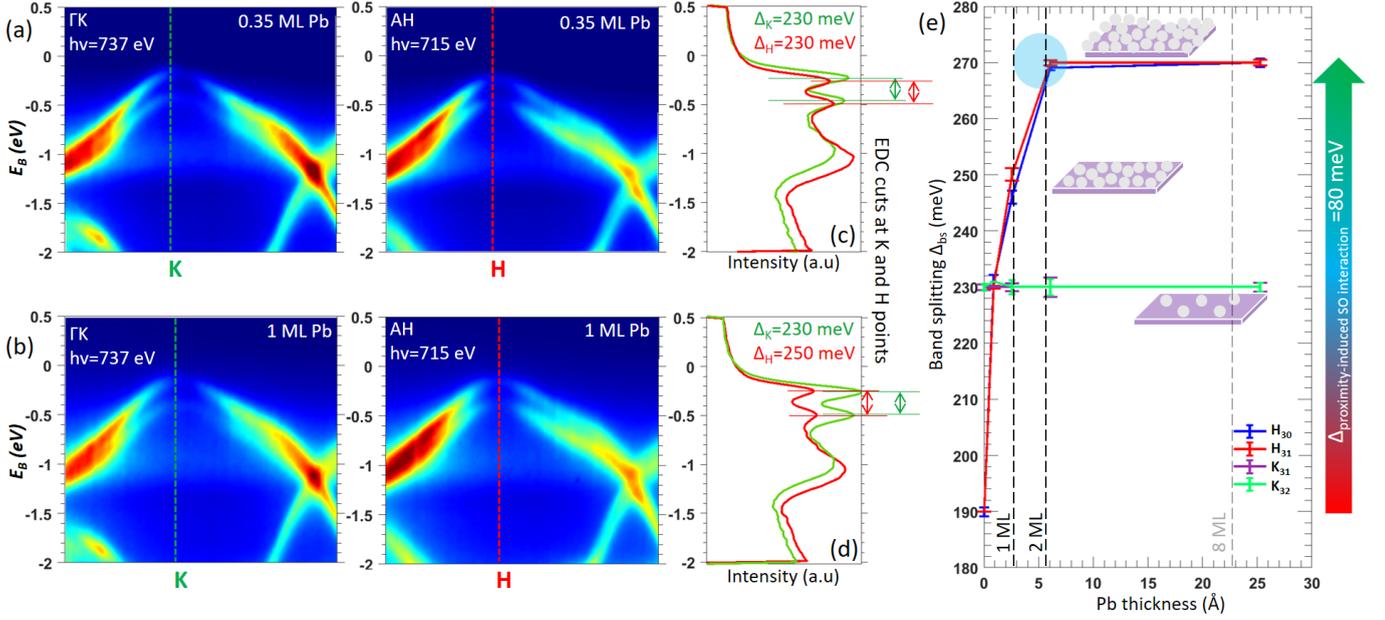

fig. 6: **Modulation of the SO band splitting at the Pb/MoSe$_2$ interface.** (a,b) ARPES images around the K (left) and H (right), corresponding to the zoom-in images in **Fig. 4 (c)**, at the indicated Pb-layer thicknesses 0.35 and 1 MLs. (c,d) The EDCs at the K and H-points on the right show the $\Delta_{bs}$ whose increase with the Pb thickness in the H point manifests the modulation of SO splitting. (e) Evolution of $\Delta_{bs}$ as a function of Pb layer's thickness at the $H_{30}$, $K_{31}$, $H_{31}$ and $K_{32}$ points, see **Fig. 4 (a)**. The blue circle represents a saturation point where $\Delta_{bs}$ reaches its maximum value of 270 meV after deposition of ~2 ML of Pb.

The radical difference in the $\Delta_{bs}$ behaviour at the K and H-points after Pb deposition can be understood by our first-principles calculations on bare MoSe$_2$ in **Fig. 2**. We have seen that at the K point $\Delta_{bs}$ is formed predominantly by the interlayer interaction, whereas at the H point $\Delta_{bs} \sim \Delta_{SO}$. Thus, the Pb layer hardly affects the bands at K, but, via the proximity of MoSe$_2$ to the high-Z atoms introducing a strong SO field, $\Delta_{bs}$ is strongly modulated at H. Our experiments on MoSe$_2$ interfaced with Pb provide the first unambiguous demonstration of the interfacial SO proximity effect on the band structure caused by heavy-Z materials. The mechanism of this effect is the hybridization of the Pb wavefunctions with the Mo-derived ones across the Pb/MoSe$_2$ interface. This hybridization is mediated by the Se-derived wavefunctions in the top layer of MoSe$_2$. Importantly, the experimental dependence of $\Delta_{bs}$ on the Pb thickness demonstrates a possibility of tuning of the SO proximity effect strength.

We have also analysed a possibility of alternative scenarios for the observed experimental results. First, the $\Delta_{bs}$ modulation might in principle have been caused by some kind of chemical reaction between Pb and MoSe$_2$. However, our XPS spectra of **Fig. 5 (a)** show no evidence for new components formed at the interface: the intensity of the Mo 3d and Pb 4f peaks only decreases and increases, respectively, during the Pb deposition without any lineshape changes. Moreover, we have spotted no changes in the ARPES dispersions, e.g. formation of new states, other than the modulation of $\Delta_{bs}$ in the SO hotspot upon the deposition of Pb. In the second scenario, the $\Delta_{bs}$ modulation might have been caused by band bending in the semiconducting MoSe$_2$ caused by the Pb overlayer via a mechanism similar to the band-gap modulation upon deposition of alkali atoms on other vdW materials [68-71]. Most susceptible to this mechanism are the out-of-plane orbitals, where a different overlap of the bonding and antibonding states in the band-bending potential changes the band gap between them. In our case, however, the ARPES data before and after the Pb deposition have not shown any changes in energy separation between the bonding Se $4p_z$ and antibonding $4p_z^*$ orbitals within the experimental errors (see Supplementary). Such an effect should be yet smaller for the Mo-derived orbitals, separated by two layers of Se atoms in the lattice. The absence of any $\Delta_{bs}$ modulation in the K-point local band gap is another reason to rule out the band-bending scenario. Thirdly, as the ultimate test for the role of the SO coupling, we have performed the same experiments upon deposition of Al, a much lighter element than Pb (see Supplementary); no changes in $\Delta_{bs}$ at either the K or H-point were detected.

# Conclusions and outlook

We have experimentally demonstrated an approach to modulate the SO interaction via the proximity effect, offering opportunities not only to pursue further fundamental research in spin physics but also develop advanced technological applications requiring a strong SO field.

Using soft-X-ray ARPES, we have provided the first, to the best of our knowledge, experimental evidence of the **k**-dependent proximity-induced modulation of the SO interaction at the interface of amorphous Pb with MoSe$_2$. This modulation has been identified through $\Delta_{bs}$ the valence band splitting of the Mo $4d_{xz}$ states in the MoSe$_2$ substrate. We have found the maximal susceptibility of $\Delta_{bs}$ to the proximity effect in the H-point of the 3D-BZ, the SO-interaction hotspot in pristine MoSe$_2$. On the other hand, the effect vanishes in the sister K-point, where $\Delta_{bs}$ is almost entirely due to the interlayer interaction. This difference assigns the observed change of $\Delta_{bs}$ to the modulation of $\Delta_{SO}$ the SO splitting due the proximity to Pb. It is caused by a transfer of the strong SO field from Pb layers to the substrate through the hybridization of Pb $sp$ states with Mo $4d$ ones, leading to formation of new proximitized hybrid states. We have carefully analysed and finally ruled out any other potential sources of the observed $\Delta_{bs}$ modulation such as a potential band-bending effect or chemical reactions at the interface.

It is worth it to note, finally, that the use of soft-X-ray ARPES has been instrumental for our observation of the **k**-dependent SO proximity effect. The large probing depth of this technique was needed to penetrate through the Pb film to MoSe$_2$, and its sharp intrinsic $k_z$ definition was essential to select the SO hotspot in the H-point of the 3D-BZ. Moreover, the amorphous character of the Pb layers eliminated potential effects of **k**-space mismatch between the proximitized wavefunctions at the Pb/MoSe$_2$ interface. The proximity-induced boost of the SO interaction increases up with the Pb thickness up to a saturation value of ~30% achieved for thicknesses above 2 monolayer, which should be sufficient for real spintronic devices.

The proximity-induced modulation of the SO interaction offers a promising route for tuning the spin-related properties of materials and tailoring them to various device applications. For instance, the heterostructure PbS/MoSe$_2$ should also allow modulation of the SO coupling in MoSe$_2$. The PbS layers here will secure the semiconductive nature of the system, allowing integration of the spin functionality into vdW transistors. Going along the route, interfacing two materials with otherwise mutually exclusive properties like superconductivity and ferromagnetism appears as the way to engineer multifunctional materials opening a wide range of applications in spintronics, quantum computing and other related fields.

# Experimental and Theoretical Methods

***ARPES measurements.*** Synchrotron-radiation ARPES experiments on a series of six commercially available 2H-MoSe$_2$ bulk crystals were performed at the SX-ARPES endstation [45] of the ADRESS beamline [72] at the Swiss Light Source (Villigen, Switzerland). The spectra were recorded in the *hv* range 700-850 eV with circularly polarised light, using a hemispherical analyzer at a combined (beamline + analyzer) energy resolution of ~100-120 meV. The samples were cleaved and measured in-situ at 14 K in a vacuum better than 2.10$^{-10}$ mbar. For each photon energy and because of charging, the Fermi level was aligned using core level spectra (Mo 3$d$) at the same *hv.* Conversion of the photoelectron kinetic energies and emission angles to the **k** values in the valence band was corrected including the photon momentum [45].

**Pb deposition.** Amorphous Pb thin layers were grown on a cooled MoSe$_2$ substrate (14 K) using a thermal Pb evaporator. The calibrated deposition rate was about 6 Å/min, i.e., it takes about 28 seconds to deposit 1 ML Pb (2.86 Å) [73].

**DFT calculations.** Electronic band structure calculations for bulk MoSe$_2$ with and without SO coupling were carried out within the density functional theory using the linear muffin-tin orbital (LMTO) method [74]. We use the fully relativistic PY LMTO computer code [75]. Self-consistent calculations of bulk 2H-MoSe$_2$ have been performed using the experimental lattice parameters.

# References

1. S. A. Wolf *et al.* Spintronics: A Spin-Based Electronics Vision for the Future. *Science* **294**, 1488–1495 (2001).

2. Awschalom, D. D. *Semiconductor Spintronics and Quantum Computation*. 311 (Springer Berlin Heidelberg, 2002).

3. A. Manchon *et al.* Current-induced spin-orbit torques in ferromagnetic and antiferromagnetic systems. *Reviews of Modern Physics* **91**, 035004 (2019).

4. Hirohata, A. *et al.* Review on spintronics: Principles and device applications. *Journal of Magnetism and Magnetic Materials* **509**, 166711 (2020).

5. Nagaosa, N., Sinova, J., Onoda, S., A. H. MacDonald & N. P. Ong. Anomalous Hall effect. *Reviews of Modern Physics* **82**, 1539–1592 (2010).

6. T. McGuire & R. Potter. Anisotropic magnetoresistance in ferromagnetic 3d alloys. *IEEE Transactions on Magnetics* **11**, 1018–1038 (1975).

7. Dyakonov, M. I. *Spin physics in semiconductors*. 439 (Springer, 2008).

8. S. D. Ganichev *et al.* Spin-galvanic effect. *Nature* **417**, 153–156 (2002).

9. K. Kato, Y. & D. Awschalom, D. Electrical Manipulation of Spins in Nonmagnetic Semiconductors. *Journal of the Physical Society of Japan* **77**, 031006 (2008).

10. Chen, W. Edelstein and inverse Edelstein effects caused by the pristine surface states of topological insulators. *Journal of Physics: Condensed Matter* **32**, 035809 (2019).

11. Shen, K., G. Vignale & R. Raimondi. Microscopic Theory of the Inverse Edelstein Effect. *Physical Review Letters* **112**, 096601 (2014).

12. H. Geng *et al.* Theory of Inverse Edelstein Effect of The Surface States of A Topological Insulator. *Scientific Reports* **7**, 3755 (2017).

13. S. Ghiasi, T., A. Kaverzin, A., J. Blah, P. & J. van Wees, B. Charge-to-Spin Conversion by the RashbaEdelstein Effect in Two-Dimensional van der Waals Heterostructures up to Room Temperature. *Nano Letters* **19**, 5959–5966 (2019).

14. Song, Q. *et al.* Observation of inverse Edelstein effect in Rashba-split 2DEG between $SrTiO_3$ and $LaAlO_3$ at room temperature. *Science Advances* **3**, 1602312 (2017).

15. Zhang, H. *et al.* Topological insulators in $Bi_2Se_3$, $Bi_2Te_3$ and $Sb_2Te_3$ with a single Dirac cone on the surface. *Nature Physics* **5**, 438–442 (2009).

16. Y. Xia *et al.* Observation of a large-gap topological-insulator class with a single Dirac cone on the surface. *Nature Physics* **5**, 398–402 (2009).

17. Yan, B. & Felser, C. Topological Materials: Weyl Semimetals. *Annual Review of Condensed Matter Physics* **8**, 337–354 (2017).

18. Wan, X., M. Turner, A., Vishwanath, A. & Y. Savrasov, S. Topological semimetal and Fermi-arc surface states in the electronic structure of pyrochlore iridates. *Physical Review B* **83**, 205101 (2011).

19. Balents, L. Weyl electrons kiss. *Physics* **4**, 36 (2011).

20. J. A. Reyes-Retana & F. Cervantes-Sodi. Spin-orbital effects in metal-dichalcogenide semiconducting monolayers. *Scientific Reports* **6**, 24093 (2016).

21. N.P. Armitage, E.J. Mele & Vishwanath, A. Weyl and Dirac semimetals in three-dimensional solids. *Reviews of Modern Physics* **90**, 015001 (2018).

22. B. M. Schröter, N. *et al.* Chiral topological semimetal with multifold band crossings and long Fermi arcs. *Nature Physics* **15**, 759–765 (2019).

23. M. M. Desjardins *et al.* Synthetic spin orbit interaction for Majorana devices. *Nature Materials* **18**, 1060–1064 (2019).

24. Das, A. *et al.* Zero-bias peaks and splitting in an AlInAs nanowire topological superconductor as a signature of Majorana fermions. *Nature Physics* **8**, 887–895 (2012).

25. Fabián Gonzalo Medina *et al.* Manipulation of Majorana bound states in proximity to a quantum ring with Rashba coupling. *Scientific Reports* **12**, 1071 (2022).

26. Sato, M. & Fujimoto, S. Topological phases of noncentrosymmetric superconductors: Edge states, Majorana fermions, and non-Abelian statistics. *Physical Review B* **79**, 094504 (2009).

27. Tanaka, Y., Mizuno, Y., Yokoyama, T., Yada, K. & Sato, M. Anomalous Andreev Bound State in Noncentrosymmetric Superconductors. *Physical Review Letters* **105**, 097002 (2010).

28. Sato, M. Majorana fermions in topological superconductors with spin-orbit interaction. *Journal of Physics: Conference Series* **391**, 012150 (2012).

29. S. M. Albrecht *et al.* Exponential protection of zero modes in Majorana islands. *Nature* **531**, 206–209 (2016).

30. S. Nadj-Perge *et al.* Observation of Majorana fermions in ferromagnetic atomic chains on a superconductor. *Science* **346**, 602–607 (2014).

31. M Shikin, A. *et al.* Induced spin orbit splitting in graphene: the role of atomic number of the intercalated metal andi/id hybridization. *New Journal of Physics* **15**, 013016 (2013).

32. D. Marchenko *et al.* Giant Rashba splitting in graphene due to hybridization with gold. *Nature Communications* **3**, 1232 (2012).

33. D. Marchenko, J. Sánchez-Barriga, M. R. Scholz, O. Rader & A. Varykhalov. Spin splitting of Dirac fermions in aligned and rotated graphene on Ir(111). *Physical Review B* **87**, 115426 (2013).

34. A. Varykhalov *et al.* Tunable Fermi level and hedgehog spin texture in gapped graphene. *Nature Communications* **6**,7610 (2015).

35. I. Klimovskikh, I. *et al.* SpinOrbit Coupling Induced Gap in Graphene on Pt(111) with Intercalated Pb Monolayer. *ACS Nano* **11**, 368–374 (2017).

36. I. Klimovskikh, I. *et al.* Reply to Comment on `SpinOrbit Coupling Induced Gap in Graphene on Pt(111) with Intercalated Pb Monolayer'. *ACS Nano* **11**, 10630–10632 (2017).

37. Dedkov, Y. & Voloshina, E. Comment on SpinOrbit Coupling Induced Gap in Graphene on Pt(111) with Intercalated Pb Monolayer. *ACS Nano* **11**, 10627–10629 (2017).

38. Avsar, A., Tan, J., Taychatanapat, T. *et al.* Spin–orbit proximity effect in graphene. *Nat Commun* **5**, 4875 (2014).

39. Lihuan Sun, Louk Rademaker, et al. Determining spin-orbit coupling in graphene by quasiparticle interference imaging. arXiv:2212.04926

40. Manzeli, S., Ovchinnikov, D., Pasquier, D., V. Yazyev, O. & Kis, A. 2D transition metal dichalcogenides. *Nature Reviews Materials* **2**, 17033 (2017).

41. Gmitra, M. & Fabian, J. Graphene on transition-metal dichalcogenides: A platform for proximity spin-orbit physics and optospintronics. *Physical Review B* **92**, 155403 (2015).

42. Gmitra, M. & Fabian, J. Proximity Effects in Bilayer Graphene on Monolayer $WSe_2$: Field-Effect Spin Valley Locking, Spin-Orbit Valve, and Spin Transistor. *Physical Review Letters* **119**, 146401 (2017).

43. Ashton, M., Paul, J., B. Sinnott, S. & G. Hennig, R. Topology-Scaling Identification of Layered Solids and Stable Exfoliated 2D Materials. *Physical Review Letters* **118**, 106101 (2017).

44. V.N. Strocov *et al.* k-resolved electronic structure of buried heterostructure and impurity systems by soft-X-ray ARPES. *Journal of Electron Spectroscopy and Related Phenomena* **236**, 1–8 (2019).

45. V. N. Strocov *et al.* Soft-X-ray ARPES facility at the ADRESS beamline of the SLS: concepts, technical realisation and scientific applications. *Journal of Synchrotron Radiation* **21**, 32–44 (2013).

46. V.N. Strocov. Intrinsic accuracy in 3-dimensional photoemission band mapping. *Journal of Electron Spectroscopy and Related Phenomena* **130**, 65–78 (2003).

47. Splendiani, A. *et al.* Emerging Photoluminescence in Monolayer $MoS_2$. *Nano Letters* **10**, 1271–1275 (2010).

48. A. Manchon, H. C. Koo, J. Nitta, S. M. Frolov & R. A. Duine. New perspectives for Rashba spin orbit coupling. *Nature Materials* **14**, 871–882 (2015).

49. Li, T. & Galli, G. Electronic Properties of $MoS_2$ Nanoparticles. *The Journal of Physical Chemistry C* **111**, 16192–16196 (2007).

50. Z. Y. Zhu, Y. C. Cheng & U. Schwingenschlögl. Giant spin-orbit-induced spin splitting in two-dimensional transition-metal dichalcogenide semiconductors. *Physical Review B* **84**, 153402 (2011).

51. B. Radisavljevic, A. Radenovic, J. Brivio, V. Giacometti & A. Kis. Single-layer $MoS_2$ transistors. *Nature Nanotechnology* **6**, 147–150 (2011).

52. Chhowalla, M. *et al.* The chemistry of two-dimensional layered transition metal dichalcogenide nanosheets. *Nature Chemistry* **5**, 263–275 (2013).

53. Hua Wang, Q., Kalantar-Zadeh, K., Kis, A., N. Coleman, J. & S. Strano, M. Electronics and optoelectronics of two-dimensional transition metal dichalcogenides. *Nature Nanotechnology* **7**, 699–712 (2012).

54. S. Lebègue, T. Björkman, M. Klintenberg, R. M. Nieminen & O. Eriksson. Two-Dimensional Materials from Data Filtering and Ab Initio-Calculations. *Physical Review X* **3**, 031002 (2013).

55. Ajayan, P., Kim, P. & Banerjee, K. Two-dimensional van der Waals materials. *Physics Today* **69**, 38–44 (2016).

56. A. Rasmussen, F. & S. Thygesen, K. Computational 2D Materials Database: Electronic Structure of Transition-Metal Dichalcogenides and Oxides. *The Journal of Physical Chemistry C* **119**, 13169–13183 (2015).

57. S. Park *et al.* Electronic band dispersion determination in azimuthally disordered transition-metal dichalcogenide monolayers. *Communications Physics* **2**, 68 (2019).

58. Bussolotti, F. *et al.* Electronic properties of atomically thin $MoS_2$ layers grown by physical vapour deposition: band structure and energy level alignment at layer/substrate interfaces. *RSC Advances* **8**, 7744–7752 (2018).

59. Alidoust, N. *et al.* Observation of monolayer valence band spin-orbit effect and induced quantum well states in $MoX_2$. *Nature Communications* **5**, 4673(2014).

60. Silva-Guillén, J., San-Jose, P. & Roldán, R. Electronic Band Structure of Transition Metal Dichalcogenides from Ab Initio and SlaterKoster Tight-Binding Model. *Applied Sciences* **6**, 284 (2016).

61. Zollner, K., E. Faria Junior, P. & Fabian, J. Strain-tunable orbital, spin-orbit, and optical properties of monolayer transition-metal dichalcogenides. *Physical Review B* **100**, 195126 (2019).

62. Cheng, Y. & Schwingenschlögl, U. $MoS_2$: A First-Principles Perspective. in *Lecture Notes in Nanoscale Science and Technology* 103–128 (Springer International Publishing, 2013).

63. D. Pescia, A.R. Law, M.T. Johnson & H.P. Hughes. Determination of observable conduction band symmetry in angle-resolved electron spectroscopies: Non-symmorphic space groups. *Solid State Communications* **56**, 809–812 (1985).

64. F. Weber *et al.* Three-dimensional Fermi surface of $2H$-$NbSe_2$ : Implications for the mechanism of charge density waves. *Physical Review B* **97**, 235122 (2018).

65. S. Tanuma, C.J. Powell & D.R. Penn. Proposed formula for electron inelastic mean free paths based on calculations for 31 materials. *Surface Science Letters* **192**, L849–L857 (1987).

66. Shinotsuka, H., Tanuma, S., J. Powell, C. & R. Penn, D. Calculations of electron inelastic mean free paths. XII. Data for 42 inorganic compounds over the 50 eV to 200 keV range with the full Penn algorithm. *Surface and Interface Analysis* **51**, 427–457 (2018).

67. Student. The Probable Error of a Mean. *Biometrika* **6**, 1 (1908).

68. Ubrig, N. *et al.* Design of van der Waals interfaces for broad-spectrum optoelectronics. *Nature Materials* **19**, 299–304 (2020).

69. N. Ehlen *et al.* Direct observation of a surface resonance state and surface band inversion control in black phosphorus. *Physical Review B* **97**, 045143 (2018).

70. Kim, J. *et al.* Observation of tunable band gap and anisotropic Dirac semimetal state in black phosphorus. *Science* **349**, 723–726 (2015).

71. G. Kremer *et al.* Ultrafast dynamics of the surface photovoltage in potassium-doped black phosphorus. *Physical Review B* **104**, 035125 (2021).

72. V. N. Strocov *et al.* High-resolution soft X-ray beamline ADRESS at the Swiss Light Source for resonant inelastic X-ray scattering and angle-resolved photoelectron spectroscopies. *Journal of Synchrotron Radiation* **17**, 631–643 (2010).

73. S. Stolyarov, V. *et al.* Superconducting Long-Range Proximity Effect through the Atomically Flat Interface of a $Bi_2Te_3$ Topological Insulator. *The Journal of Physical Chemistry Letters* **12**, 9068–9075 (2021).

74. O. Krogh Andersen. Linear methods in band theory. *Physical Review B* **12**, 3060–3083 (1975).

75. Antonov, V., Yaresko, A. & Harmon, B. *Electronic Structure and Magneto-optical Properties of Solids*. 545 (Springer-Verlag New York Inc.).

# k-dependent proximity-induced modulation of spin-orbit interaction in MoSe$_2$ interfaced with amorphous Pb
# (Supplementary materials)

*Fatima Alarab* [1], *Ján Minár* [2], *Procopios Constantinou* [1], *Dhani Nafday*[3], *Thorsten Schmitt* [1], *Xiaoqiang Wang* [1] *and Vladimir N. Strocov* [1]

[1]Swiss Light Source, Paul Scherrer Institute, 5232 Villigen-PSI, Switzerland
[2]New Technologies Research Centre, University of West Bohemia, 301 00 Plzeň, Czech Republic
[3]Asia Pacific Center for Theoretical Physics, Pohang, Gyeongbuk, 37673, South Korea

1. Pb deposition on MoSe$_2$: Background intensity and energy gap between the Se $p_z/p_z^*$ states as a function of Pb thickness

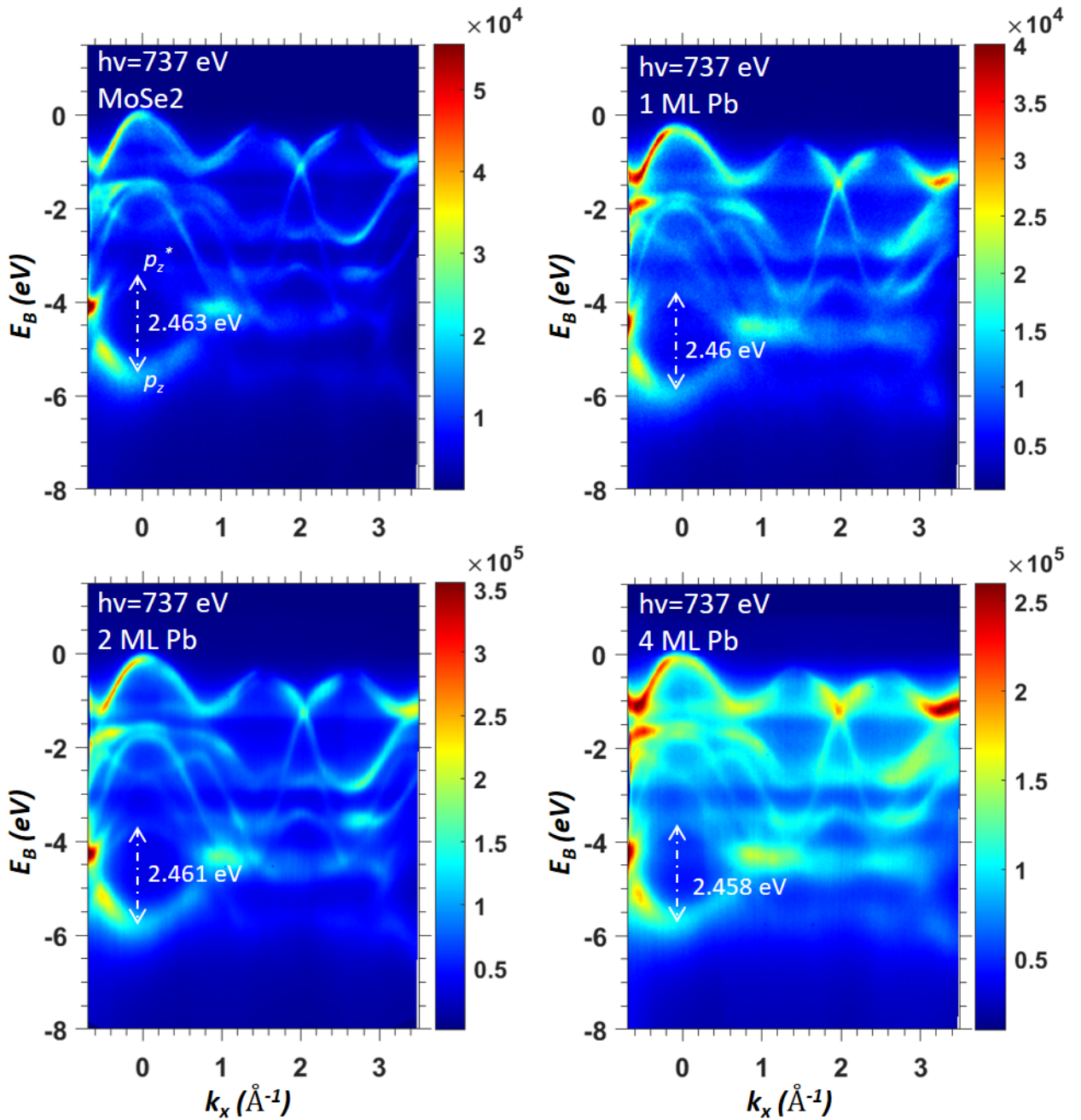

**SFigure 1**: ARPES intensity images measured at *hv*=737 eV for pristine MoSe$_2$ and through a series of Pb overlayer thicknesses (0, 1, 2, and 4 MLs). The incoherent background, coming from the amorphous Pb overlayers, progressively increases with the Pb thickness, without any noticeable change in the band dispersions. The gap between the Se bonding ($p_z$) and antibonding ($p_z^*$) states at the Γ point ($k_x$=0) is indicated. Its magnitude does not show any statistically significant variation upon the Pb deposition.

## 2. Al deposition on MoSe$_2$: Band splitting in the K and H points

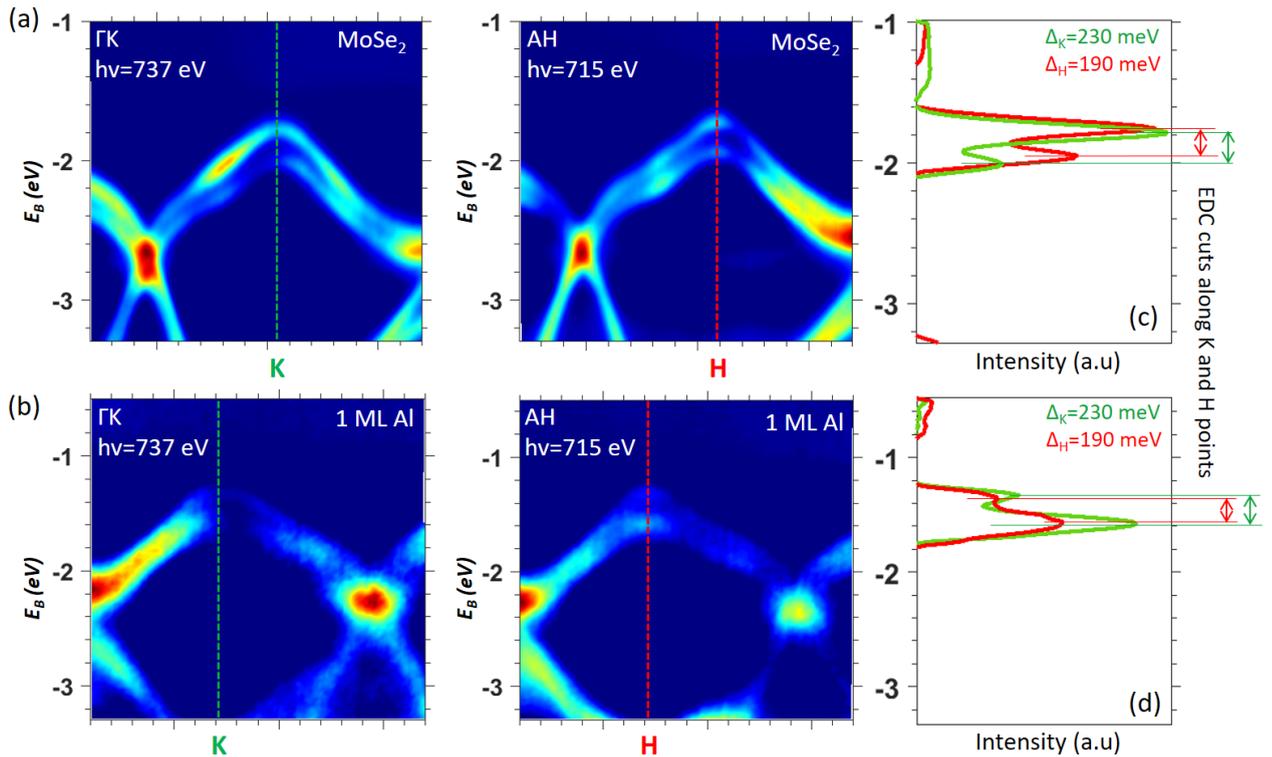

**SFigure 2**: (a) ARPES images of bulk MoSe$_2$ and (b) after the deposition of 1 ML Al, around the K (left) and H point (right). (c,d) The corresponding EDC cuts. The Al overlayer does not change the band splitting at either K or H points.

## 3. XPS core level intensity evolution upon Pb deposition

Based on the decreasing intensity of the Mo 3$d$ core levels relative to the Pb 4$f$ during Pb deposition, see **Fig. 5 (a)**, we can calculate the average thickness of the grown ovelayers using $I=I_0\ e^{-d/\lambda}$, where $I$ is the core level intensity (here we focus on Mo 3$d$ and Se 3$d$), $\lambda$ is the mean free path of the electrons ejected throughout the Pb layers ($\lambda$=19.5 Å at photon energy of 1000 eV [65,66]), $I_0$ is the intensity of the core level for the pure sample, and $d$ is the Pb overlayer thickness. The correspondence between the deposition time and overlayer thickness presented in **Fig. 5 (b)** shows a linear behaviour. The two curves obtained from the intensity attenuation of the Mo 3$d$ and Se 3$d$ core levels show similar linear behaviour, which can be splitted in two windows. The first one that starts from 0 to 50 seconds of Pb deposition, shows a linear increase of Pb layer's thickness with deposition time. Yet, the second part (between 50 to 100 seconds) even though it also shows a linear behaviour, the slope of the curve is higher. At 50 sec, the surface of MoSe$_2$ is fully covered with Pb atoms, which therefore meaningfully changes the spectral weight of the core levels coming from the substrate before and after this point. At a deposition time below 50 sec, the surface of MoSe$_2$ is partially covered with Pb atoms. In this case, the Mo/Se related photoelectrons are not entirely attenuated by Pb atoms deposited on top of the substrate, and therefore the following relation $I=I_0\ e^{-d/\lambda}$ is not fully appropriate, missing an additional non-attenuated component. After completely covering the surface with Pb atoms, the above attenuation rule for the core level intensity becomes fairly accurate.

We also argue that the difference of the obtained Pb thickness values using Mo 3$d$ and Se 3$d$ core levels can be attributed to the surface termination of the bulk substrate (Se atoms on top and Mo atoms below) and an atomic disorder at the interface. As Se atoms are likely to be closer to Pb than Mo, the photoelectrons scatter less and therefore lead to smaller attenuation compared to Mo atoms.